\documentclass[allclo]{FBSart}
\usepackage{amsfonts}
\usepackage{amssymb}

\usepackage{amsbsy}

\usepackage[dvips]{graphicx,color}

\begin{document}

\title{Poincar{\'e} Invariant Three-Body Scattering}

\author{Ch.~Elster$^{(a)}$, T.~Lin$^{(a)}$, W.N. Polyzou$^{(b)}$, 
W.~Gl\"ockle$^{(c)}$}
\institute{(a)
Institute of Nuclear and Particle Physics,  and
Department of Physics and Astronomy,  Ohio University,
Athens, OH 45701, USA \\
(b) Department of Physics and Astronomy, The University of Iowa, Iowa
City, IA 52242, USA\\
(c) Institute for Theoretical Physics II, Ruhr-University Bochum,
D-44780 Bochum, Germany}

\runningauthor{Ch. Elster}
\runningtitle{Poincar\'e Invariant Three-Body Scattering}
\sloppy

\maketitle

\begin{abstract}
Relativistic Faddeev equations for three-body scattering are solved
at arbitrary energies in terms of momentum vectors without employing
a partial wave decomposition. Relativistic invariance is incorporated
withing the framework of Poincar\'e invariant quantum mechanics.
Based on a Malfliet-Tjon interaction, observables for elastic and
breakup scattering are calculated and compared to non-relativistic
ones.
\end{abstract}

A consistent treatment of intermediate energy reactions requires a
Poincar\'e symmetric quantum theory \cite{Wigner39}. 
In addition, the
standard partial wave decomposition, successfully applied below the
pion-production threshold~\cite{wgphysrep}, is no longer an adequate
numerical scheme due to the proliferation of the number of partial
waves.  Thus, the intermediate energy regime is a new territory for
few-body calculations, which waits to be explored.

This work addresses two aspects in this list of challenges: exact
Poincar\'e invariance and calculations using vector
variables instead of partial waves. In Ref.~\cite{Liu:2004tv} the
non-relativistic Faddeev equations were
solved directly as function of vector variables for scattering at
intermediate energies. A key advantage of this formulation lies in its
applicability at higher energies, 
where the number of partial waves proliferates.
The Faddeev equation, based on a Poincar{\'e} invariant mass
operator, has been formulated in detail in~\cite{Lin:2007ck}
and has both kinematical and dynamical
differences with respect to the corresponding non-relativistic
equation.

The formulation of the theory is given in a representation of
Poincar\'e invariant quantum mechanics where the interactions are
invariant with respect to kinematic translations and rotations
\cite{Coester65}.  The model Hilbert space is a three-nucleon Hilbert
space (thus not allowing for absorptive processes).  The method used
to introduce the NN interactions in the unitary
representation of the Poincar\'e group allows to input of e.g. 
high-precision NN interactions in a way
that reproduces the measured two-body observables. However in this
study we use a simpler interaction consisting of a superposition of
an attractive and a
repulsive Yukawa interaction with parameters chosen 
such that a bound state at $E_d$~=~-2.23~MeV is
supported~\cite{Lin:2007ck}.
Poincar\'e invariance and $S$-matrix
cluster
properties dictate how the two-body interactions must be embedded in
the three-body dynamical generators.  Scattering observables are
calculated using Faddeev equations formulated with the mass Casimir
operator (rest Hamiltonian) constructed from these generators.

\begin{figure}[t]
\begin{center}
\includegraphics*[width=6.0cm,height=4.2cm]{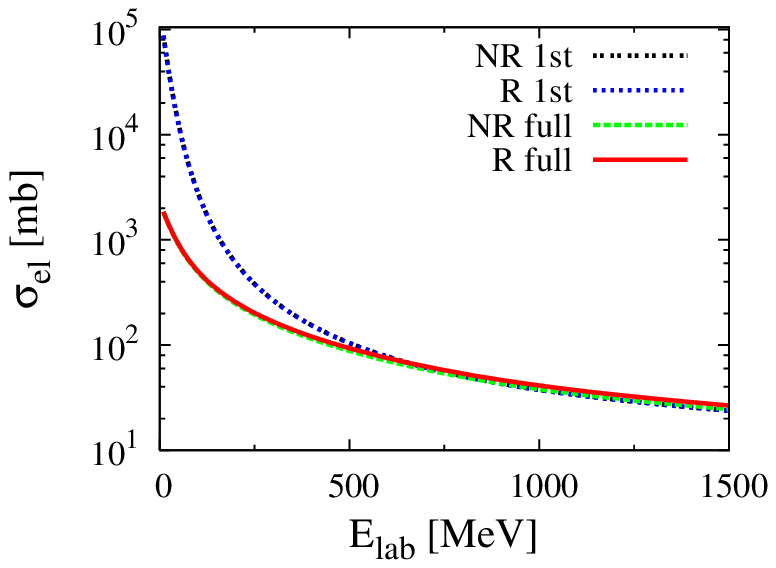}
\includegraphics*[width=6.0cm,height=4.2cm]{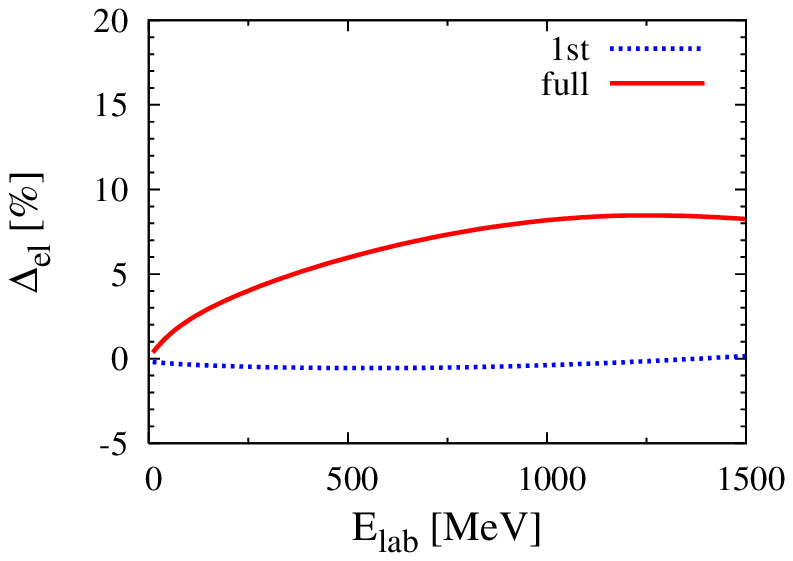}
\includegraphics*[width=6.0cm,height=4.2cm]{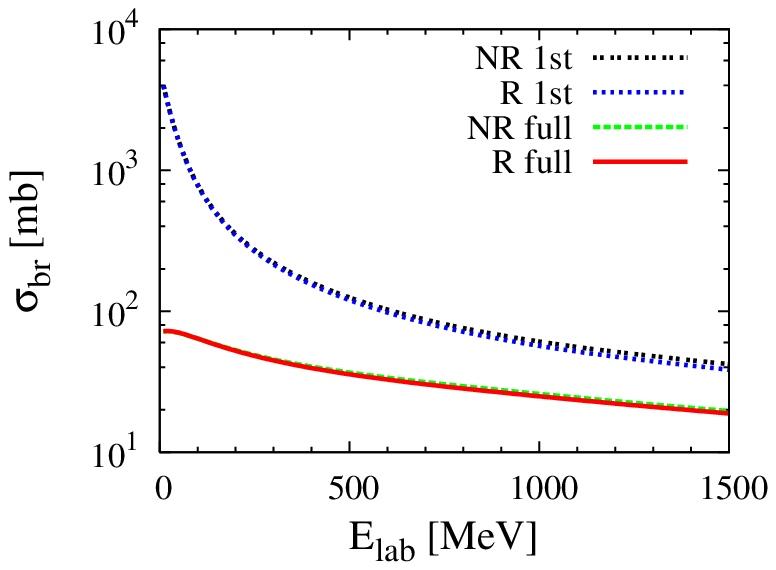}
\includegraphics*[width=6.0cm,height=4.2cm]{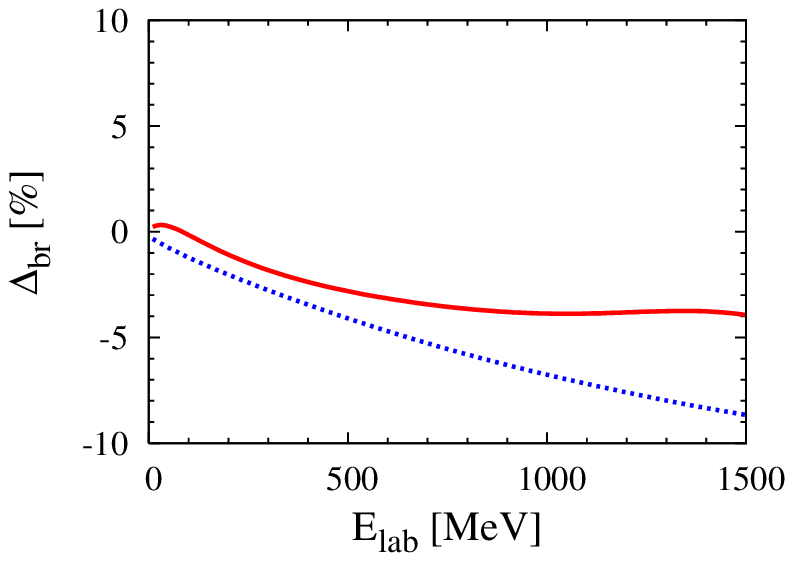}
\end{center}
\vspace{-2mm}
\caption{The total elastic c.m. cross section for elastic 
(top left) and for breakup scattering (bottom left)
calculated from a Malfliet-Tjon type potential as function of
the projectile kinetic laboratory energy. The labels `R' (`NR') stand
for relativistic (non-relativistic) calculations. The Faddeev
calculations in the first order in $t$ are marked with `1st', the
converged full Faddeev calculations with `full'. To show the
difference, the percentage difference between the relativistic and
corresponding non-relativistic calculations are displayed on the
right.}
\label{fig1}
\end{figure}

To obtain a valid estimate of the size of relativistic effects, it is
important that the interactions employed in the relativistic and
non-relativistic calculations are phase-shift equivalent. We follow
here the suggestion by Coester, Piper, and Serduke (CPS) in constructing a 
phase equivalent interaction from a non-relativistic 2N
interaction~\cite{CPS} by adding the interaction to the square of the
mass operator.
In this CPS method the relativistic interaction can not be analytically
calculated from the non-relativistic one. However, there is a simple
analytic connection between the relativistic and non-relativistic two-body
t-matrices
\begin{equation}
\nonumber
t_{re}({\bf p},{\bf p}';2E^{rel}_p) = 
\frac{2m}{\sqrt{m^2+p^2}+\sqrt{m^2+p'^2}} \; t_{nr}({\bf p},{\bf
p}';2E^{nr}_p),  
\label{cps}
\end{equation}
where $2E^{rel}_p = 2\sqrt{m^2+p^2}$ and $2E^{nr}_p =
\frac{p^2}{m}+2m$.
This relativistic two-body t-matrix 
$t_{re}({\bf p},{\bf p}';2E^{rel}_p)$ is
scattering equivalent to the non-relativistic one at the same
relative momentum {\bf p}~\cite{Keister:2005eq}. 
This t-matrix serves then as input to
obtain the Poincar{\'e} invariant transition amplitude of the 2N
subsystem embedded in the three-particle Hilbert space via a 
first resolvent method as layed out in Ref.~\cite{Lin:2007ck}.

By construction, differences in the relativistic and non-relativistic
calculations first appear in the three-body calculations.  
Those differences are in the choice of kinematic variables (Jacobi
momenta are constructed using Lorentz boosts rather then Galilean
boosts) and in the embedding of the two-body interactions in the
three-body problem, which is a consequence of the non-linear relation
between the two and three-body mass operators. These differences
modify the permutation operators and the off-shell properties of the
kernel of the Faddeev equations~\cite{Lin:2008sy}.

\begin{figure}[t]
\includegraphics*[width=13.2cm]{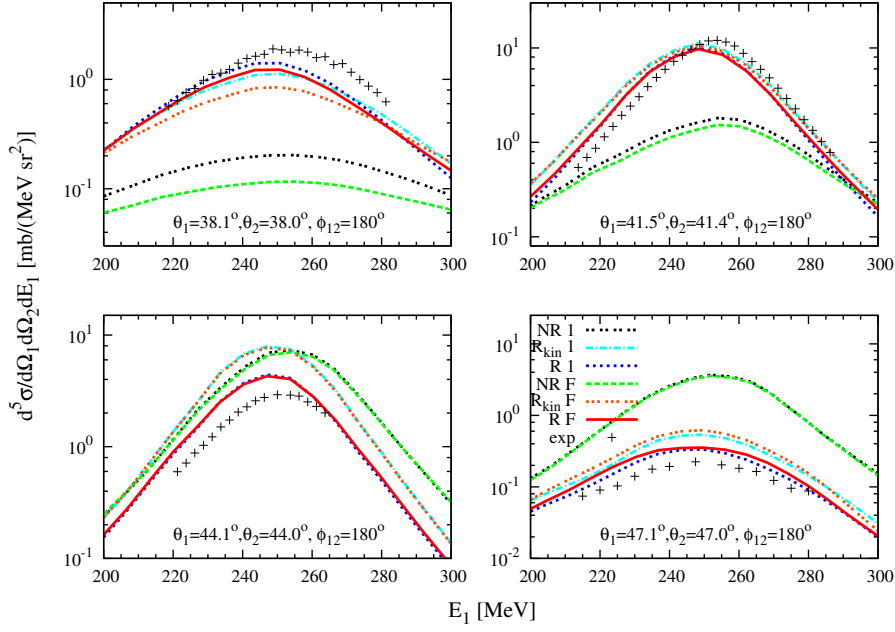}
\caption{The exclusive differential cross section for the reaction 
$^2$H(p,2p)n at 508~MeV laboratory projectile energy for different
proton angle pairs $\theta_1-\theta_2$ symmetric around the beam axis
as function of the laboratory kinetic energy of one of the outgoing
protons. The meaning of the curves are the same as in
Fig.~\ref{fig1}, except that here `1' denotes the 1st order
Faddeev calculation, `F' the fully converged one. In the curves
labeled $R_{\rm kin}$ only relativistic kinematics is taken into
account. The data are taken from Ref.~\protect\cite{Punjabi:1988hn}.
} 
\label{fig2}
\end{figure}

In Fig.~\ref{fig1} the total cross sections for elastic and breakup
cross sections are displayed as function of the projectile kinetic
energy up to 1.5~GeV obtained from our fully converged relativistic
Faddeev calculation as well as the one obtained from the first-order
term, $T^{1st}=tP$, with $P$ being the permutation operator for three
identical particles.
It is obvious that, especially for energies below 300~MeV, the
contribution of rescattering terms is huge. However, for extracting
the size of relativistic effect, it is more useful to consider 
the relative difference between the relativistic and non-relativistic
calculations. In first order, there is
essentially no effect in the total elastic cross section, 
which is consistent with the observation that
when the relativistic two-body $t$-matrix is constructed to be
phase-shift equivalent to the non-relativistic one.
Making the same
comparison with fully converged Faddeev calculations indicates that
relativistic effects in the three-body problem increase the total
cross section for elastic scattering with increasing energy, whereas it
is slightly reduced in the total 
breakup cross section.

Considering exclusive breakup reactions, differences between a
relativistic and non-relativistic calculation can be more pronounced
and strongly depend on the specific
configuration. Though our two-body force is simple, we compare 
to a $^2$H(p,2p)n experiment at
508~MeV~\cite{Punjabi:1988hn} to see if our calculation captures
essential features of the measurement. Differences in the predictions of
our relativistic and non-relativistic calculations 
are very pronounced at this energy as can be seen in Fig.~\ref{fig2},
which shows selected angle pairs $\theta_1-\theta_2$ from
Ref.~\cite{Punjabi:1988hn}, which are symmetric around the beam axis. 
The cross section is plotted against the laboratory kinetic energy of
one of the outgoing protons. It is interesting to observe that for
smaller angle pairs the relativistic cross sections (RF) are
considerably larger than the non-relativistic ones (NRF). For larger
angle pairs the situation reverses. It is further noteworthy, that in
the configurations of Fig.~\ref{fig2}, which are close to quasi-free,
rescattering effects (or equivalently higher order contributions of the
Faddeev multiple scattering series) are very small. To show that
peak-positions are given by kinematics, we added a curve labeled
`R$_{kin}$' to the figures, which stands for a non-relativistic
calculation in which only kinematics and phase space factors are
replaced by the relativistic ones.
We want to note that the above comparisons 
do not involve a non-relativistic limit, instead relativistic and
non-relativistic three-body calculations with interactions that are fit
to the same two-body data are compared. All of the differences are due
to the different ways two-body dynamics is incorporated in the three-body
problem.



\end{document}